\documentstyle[11pt,newpasp,twoside,epsf]{article}

\input{psfig}
\begin{document}

\title{Complex wind dynamics and ionization structure in symbiotic binaries}

\author{Rolf Walder\altaffilmark{1}}
\affil{Institute of Astronomy, ETH Z\"urich, Switzerland}

\author{Doris Folini\altaffilmark{2}}
\affil{Seminar of Applied Mathematics, ETH Z\"urich, Switzerland}

\altaffiltext{1}{Web-address:
                 http://www.astro.phys.ethz.ch/staff/walder/walder.html}
\altaffiltext{2}{Office currently at: Institute of Astronomy, 
                ETH Zentrum, 8092 Z\"{u}rich, Switzerland. \newline
                Web-address:
                 http://www.astro.phys.ethz.ch/staff/folini/folini.html}

\begin{abstract}
Aspects of the wind-dynamics in symbiotic binaries, colliding winds
and accretion, are reviewed. Inconsistencies between theory and
observations of the hot star wind are discussed. If the hot star wind
were governed by CAK theory, nearly all symbiotics would be colliding
wind binaries. For the case of colliding winds, 3D hydrodynamical
simulations reveal that the matter distribution is spirally
shaped. Shock confined high-density shells as well as huge voids are
found even in the immediate neighborhood of the stars. Synthetic
spectra computed on the basis of different 3D hydrodynamical models
suggest observational discrimination between them to be possible.
Colliding wind models also provide a link between symbiotics and
planetary nebulae. Accretion during some time is a necessary condition
for symbiotics to exist. However, there is no proof of whether
currently accreting systems show the symbiotic phenomenon. Existing
accretion models are inconsistent amongst each other, predicting
either extended disks or small, high-density accretion wakes.
Synthetic spectra allowing to discriminate between two models do not
yet exist.
\end{abstract}



\section{Introduction}
Symbiotics have a complex dynamical behavior. Observations show
variability on time scales of seconds to probably thousands of
years. We know of bipolar outflows and jet-like features. Radio
observations reveal complex and sometimes time-dependent structures of
the circumstellar nebula. IR-emission often shows the presence of
dust. From optical and UV spectra we take that the nebula has
different velocity and density regimes. Many symbiotics are X-ray
bright, revealing the presence of a hot ($\ga 10^7$~K) plasma. Some
systems are quiet at the moment, from others we know that they
underwent novae or smaller outbursts.

Most of these phenomena are directly or indirectly related to the wind
dynamics of symbiotic systems. However, the very number of involved
physical processes and the large range of spatial and temporal scales
prevented a consistent, quantitative model of symbiotics so far. But
however, many successful attempts towards such a model. They form the
basis of this review. In Section~2 we discuss discrepancies between
observations and CAK-theory for winds from the hot component. This
question is decisive for whether symbiotics are colliding wind or
accreting systems. In Section~3, colliding wind models, their link to
planetary nebulae, and their spectral response are presented.
Accretion models are discussed in Section~4.  Finally, a summary is
given in Section~5.
\section{To what degree can photons from the hot primary drive matter?}
\label{sec:cw_contra_accretion}
It is commonly accepted that the hot components in symbiotics are post
AGB-stars which are reborn. There may be some exceptions, where the
hot component is a neutron star or an accreting main sequence star. In
this review, these exceptions as well as recurrent novae are
explicitly excluded. For bringing back a white dwarf from its cooling
track to a post AGB state accretion is an absolutely necessary
condition. Accretion is also the basis for novae to occur and for
shell-flashes proposed to explain symbiotic outbursts. On the other
hand, we know of some symbiotics to be colliding wind systems. In this
Section we discuss whether winds can be driven from the hot star and
whether they can prevent accretion.

M\"{u}rset et al. (1991) locate hot stars of symbiotics at the same
place in the HR-diagram as central stars of planetary nebulae (CSPNe),
when excluding the above mentioned exceptions. Hot primaries of
symbiotics have temperatures above 60'000~K, many above~100'000~K, and
luminosities between 10 and $10^5$~L$_{\odot}$ with many of them
around 1000~L$_{\odot}$. Since CSPNe often loose mass, we expect fast,
radiatively driven winds from the hot component in symbiotics as well.
\subsubsection{Theory}
For radiatively (line-)driven winds CAK theory predicts a
momentum-luminosity relation (see e.g. Kudritzki 1998) of
\vspace{-0.1cm}
\begin{equation}
\dot{M} v_{\infty}  \propto R_{*}^{-0.5} L_{*}^{1/\alpha} 
\left[ M_{*} \left( 1- \Gamma \right) \right]^{3/2 - 1/\alpha} .
\label{eq:rad_driven.winds_theory}
\end{equation}
$\Gamma$ is the Eddington factor. $\alpha$ reflects the distribution
function of the oscillator strength of all lines involved in the wind
driving. Of course, $\alpha$ is a key parameter. In many cases, it is
close to 2/3, suggesting that the expression in the square brackets
has potentially no influence. As we will see, more attention should be
paid to the evaluation of photon-matter interaction for the case of
symbiotics.
\subsubsection{Observations} 
Comparing with observations (however, with not too hot stars with
radii bigger than 0.5 R$_{\odot}$), Kudritzki (1998) suggests the fits
\begin{equation}
\dot{M} v_{\infty}  = 10^A 
                      \sqrt{ R_{\odot}/R_{*}} 
                      \left( L_{*} / L_{\odot} \right)^x,
\hspace{1.5cm} v_{\infty}  =  B   v^{esc},
\label{eq:rad_driven.winds_fits}
\end{equation}
with $3 \le B \le 3.5$ and $x=1.5$ for CSPNe. For O-supergiants A is
20.65 and one expects a similar value for CSPNe. In hotter CSPNe, the
observational detection of winds is demanding. Based on IUE high
resolution spectra, Patriarchi \& Perinotto (1991) found that nearly
all CSPNe with $\log r = \log (R_{*}/ R_{\odot}) > - 1$ and some with
$\log r< -1$ show P-Cygni profiles. This revised an earlier study
based on IUE low resolution spectra (Ceruti-Sola \& Perinotto 1985),
where the limit with $\log r > -0.5$ was substantially higher. Based
on HST/GHRS observation of the CIV 155.0 nm doublet, Patriarchi \&
Perinotto (1996) report that K1-16 has a wind with 3800~km/s and a
mass loss rate as low as $\dot{\mbox{M}} < 2 \cdot
10^{-11}$~M$_{\odot}$/y. So far, this is one of the fastest winds of
CSPNe ever measured.
\subsubsection{Symbiotics}
As long as the stars are not too hot, CAK-theory fits well with the
observed evolution of winds after the outburst of symbiotic novae
(Vogel \& Nussbaumer 1994; Schmid, this volume). But there are severe
discrepancies between CAK-theory and observations for systems having a
small, hot primary.

For most symbiotics we have no observational evidence for a wind from
the hot primary; see e.g. Dumm et al. (2000) for a discussion of the
case of RW~Hya. In AG~Peg, Nussbaumer, Schmutz, \& Vogel (1995)
observe P-Cygni profiles indicating a wind with 900 km/s. Schmutz
(1996) derives from spectra taken in 1970 R$_{*} = 0.5$~R$_{\odot}$, L
$=1600$~L$_{\odot}$, v$_{\infty} = 700$~km/s, $\dot{\mbox{M}} =
10^{-6.7}$ M$_{\odot}$/y. For spectra taken in 1994 he derives R$_{*}
= 0.06$~R$_{\odot}$, L $\ga 500$~L$_{\odot}$, v$_{\infty} = 950$~km/s,
$\dot{\mbox{M}} = 10^{-6.7}$~M$_{\odot}$/y. From a theoretical point
of view, it is hard to understand why the mass loss and the velocity
from a shrinking star with decreasing luminosity stay
constant. According to CAK-theory ($B=3, A=20, x=1.5$), the wind
parameters corresponding to the same temperatures and luminosities
would be v$_{\infty} = 1700$~km/s, $\dot{\mbox{M}} = 8.4 \cdot
10^{-10}$~M$_{\odot}$/y (1970) and v$_{\infty} = 4900$~km/s,
$\dot{\mbox{M}} = 1.5 \cdot 10^{-10}$~M$_{\odot}$/y (1994). Similarly
in EG~And. Vogel (1993) derives v$_{\infty} = 500$~km/s and
$\dot{\mbox{M}}= 2 \cdot 10^{-9}$~M$_{\odot}$/y. The CAK-values
corresponding to T$_{*} = 70'000$~K and L$_{*} = 15$~L$_{\odot}$
(M\"{u}rset et al. 1991) are $7000$~km/s and $\dot{\mbox{M}}=
10^{-13}$~M$_{\odot}$/y.

Thus, either luminosities and temperatures, or wind velocities and
mass loss rates derived from observations are wrong, or CAK-theory
breaks down for the winds from primaries in symbiotics. One systematic
problem in the derivation of luminosities and temperatures by
M\"{u}rset et al. (1991) and of the mass loss rate in EG~And by Vogel
(1993) is that a spherically symmetric mass distribution around the
red star is assumed. Both, colliding wind and accretion models,
however, predict that this is by far not the case. With regard to CAK,
Springmann \& Pauldrach (1992) note that in very rarefied winds the
metals decouple from the bulk of the mass. This may lead to a much
lower outflow velocity or even to a fall-back of hydrogen and
helium. Porter \& Skouza (1999), Porter (this volume), and Krti\u{c}ka
\& Kub\'{a}t (this volume) discuss the idea in more detail. From the
investigation of Gayley (1995) follows that stars with low
Eddington-factors ($\la 5 \cdot 10^{-4}$) are no longer able to drive
winds\footnote{We thank Stan Owocki for pointing this out to us.}.
However, most of the nowadays accepted values for symbiotics lie above
this limit.
\subsubsection{Colliding wind models against accretion models}
Let us assume for a moment that
relation~\ref{eq:rad_driven.winds_fits} is indeed valid for winds from
the hot component in symbiotics. Then, assuming typical RGB- or
AGB-winds respectively, it can be estimated that {\it all systems}
with hot components having luminosities above 10~L$_{\odot}$ are
colliding wind systems. Even with very low mass loss rates, high speed
winds have enough momentum to prevent circumstellar matter from
falling onto the star. The question remains open whether this is
true.
\section{Colliding winds}
\label{sec:collwind}
Three dimensional computer models for colliding winds, even for
comparatively simple physics, are still very demanding and need a lot
of computer time. For symbiotics, only a very few have been presented,
to our knowledge all by the Z\"{u}rich group (Nussbaumer \& Walder
1993, Walder 1995a, Walder 1995b, Walder 1998). We briefly review
their results and add new ones from work in progress.

\subsection{Hydrodynamics: Carving, shaping and pushing}
\label{subsec:hydro_models}
We discuss colliding winds at the example of three different 3D
hydrodynamical models. In all models, typical for S-type symbiotics,
we assume an orbital period of two years,
1.4~M$_{\odot}$\thinspace\thinspace for the cool and
0.6~M$_{\odot}$\thinspace\thinspace for the hot star, 20 km/s for the
cool star wind and 1000 km/s for the hot star wind.  The three models
differ in the mass loss rates for which we adopt $\dot{\mbox{M}}_{c} =
3.14 \times 10^{-7}$~M$_{\odot}$/y, $\dot{\mbox{M}}_{h} = 1.  \times
10^{-9}$~M$_{\odot}$/y\thinspace\thinspace (model {\it weak}),
$\dot{\mbox{M}}_{c} = 1 \times 10^{-7}$~M$_{\odot}$/y,
$\dot{\mbox{M}}_{h} = 2 \times
10^{-9}$~M$_{\odot}$/y\thinspace\thinspace (model {\it medium}), and
$\dot{\mbox{M}}_{c} = 3.14 \times 10^{-8}$~M$_{\odot}$/y,
$\dot{\mbox{M}}_{h} = 4 \times 10^{-9}$~M$_{\odot}$/y \thinspace
\thinspace(model {\it strong}). Consequently, the ratio of the
momentum flux of the fast wind to that of the slow wind ranges from
1/6 ({\it weak}), over 1 ({\it medium}), to 6 ({\it strong}). For
simplicity, we assume that both winds have reached their terminal
velocity, and we neglect radiative forces and gravity. This assumption
is critical for close systems where the wind-wind interaction zone and
even the hot component itself may be located well within the
acceleration region of the wind from the red star. Rotation of the
stars is also neglected, which is another critical assumption, in
particular for AGB stars.

\begin{figure}[tbp]
%
\centerline{
   \plotfiddle{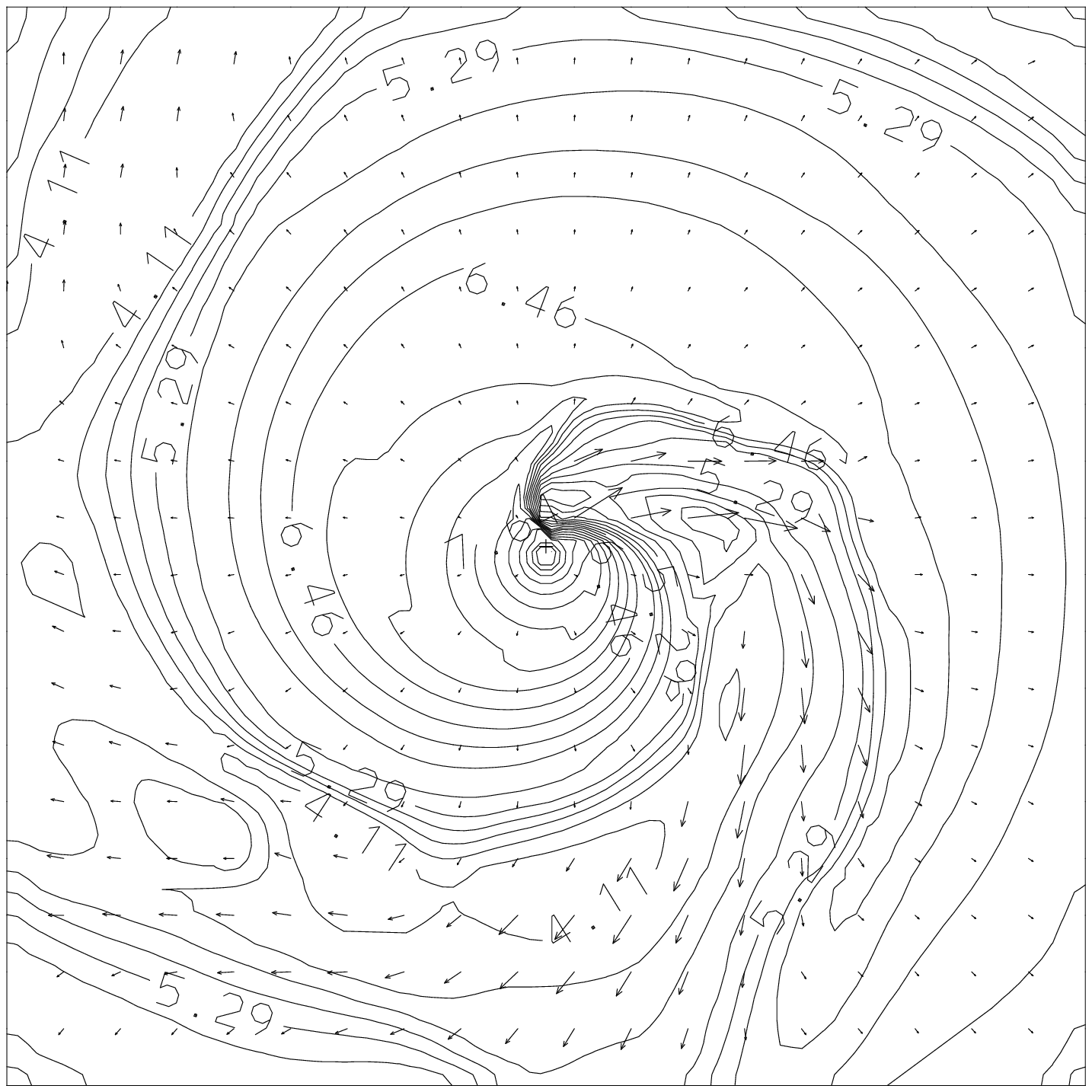}{4.4cm}{0}{30}{30}{-23}{-65}
   \plotfiddle{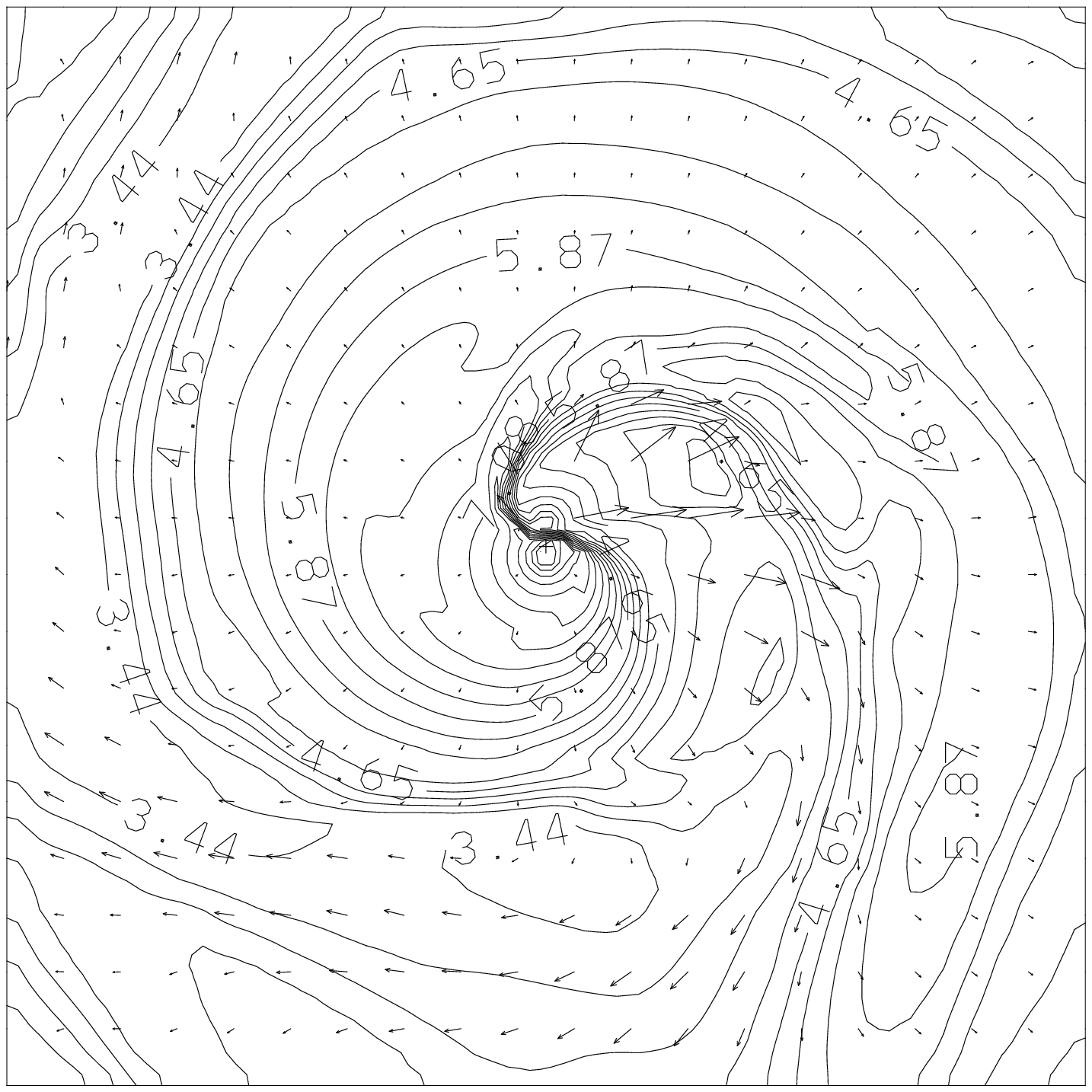}{4.4cm}{0}{30}{30}{-281}{-65}
   \plotfiddle{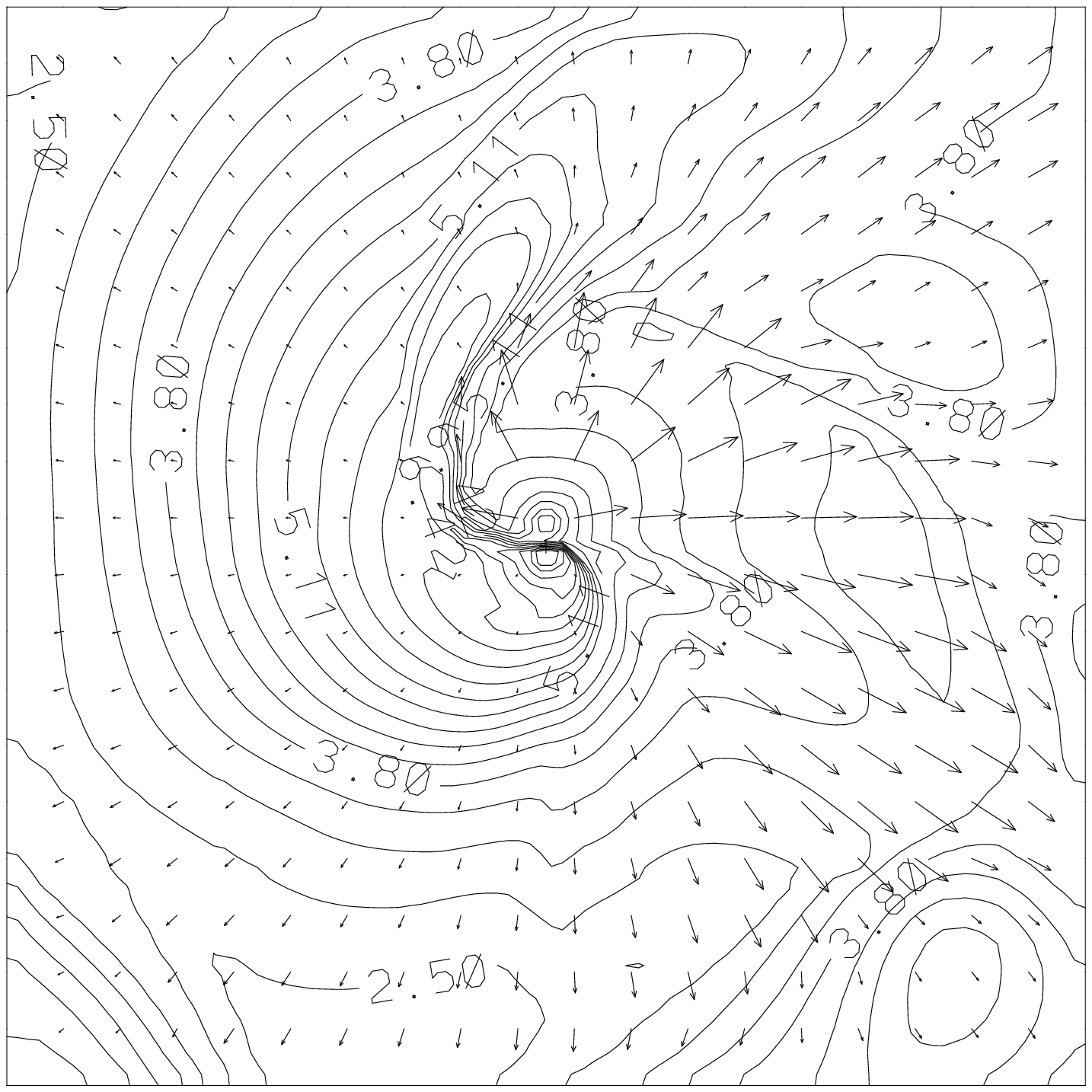}{4.4cm}{0}{30}{30}{-539}{-65}
           }
\centerline{
   \plotfiddle{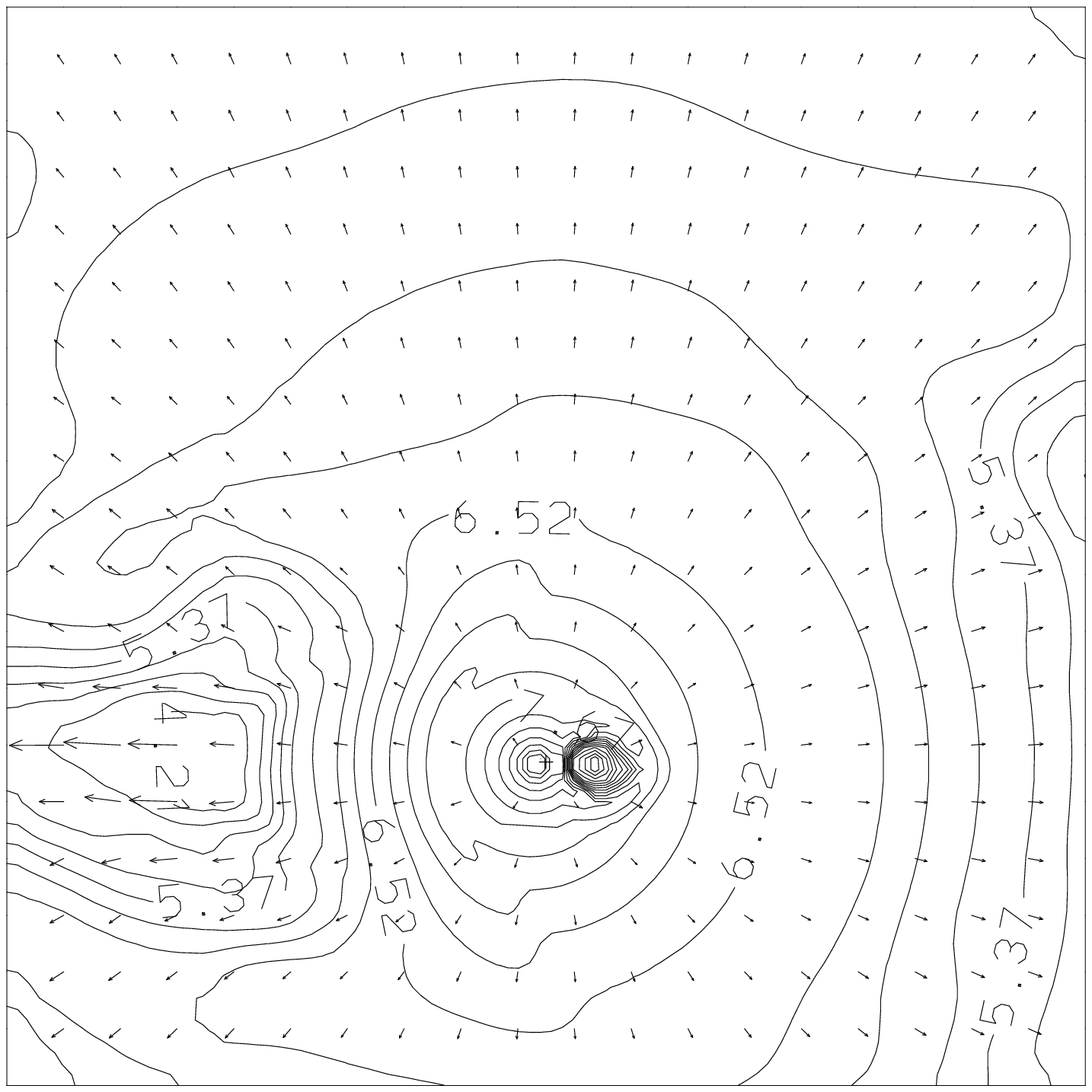}{4.4cm}{0}{30}{30}{-23}{-65}
   \plotfiddle{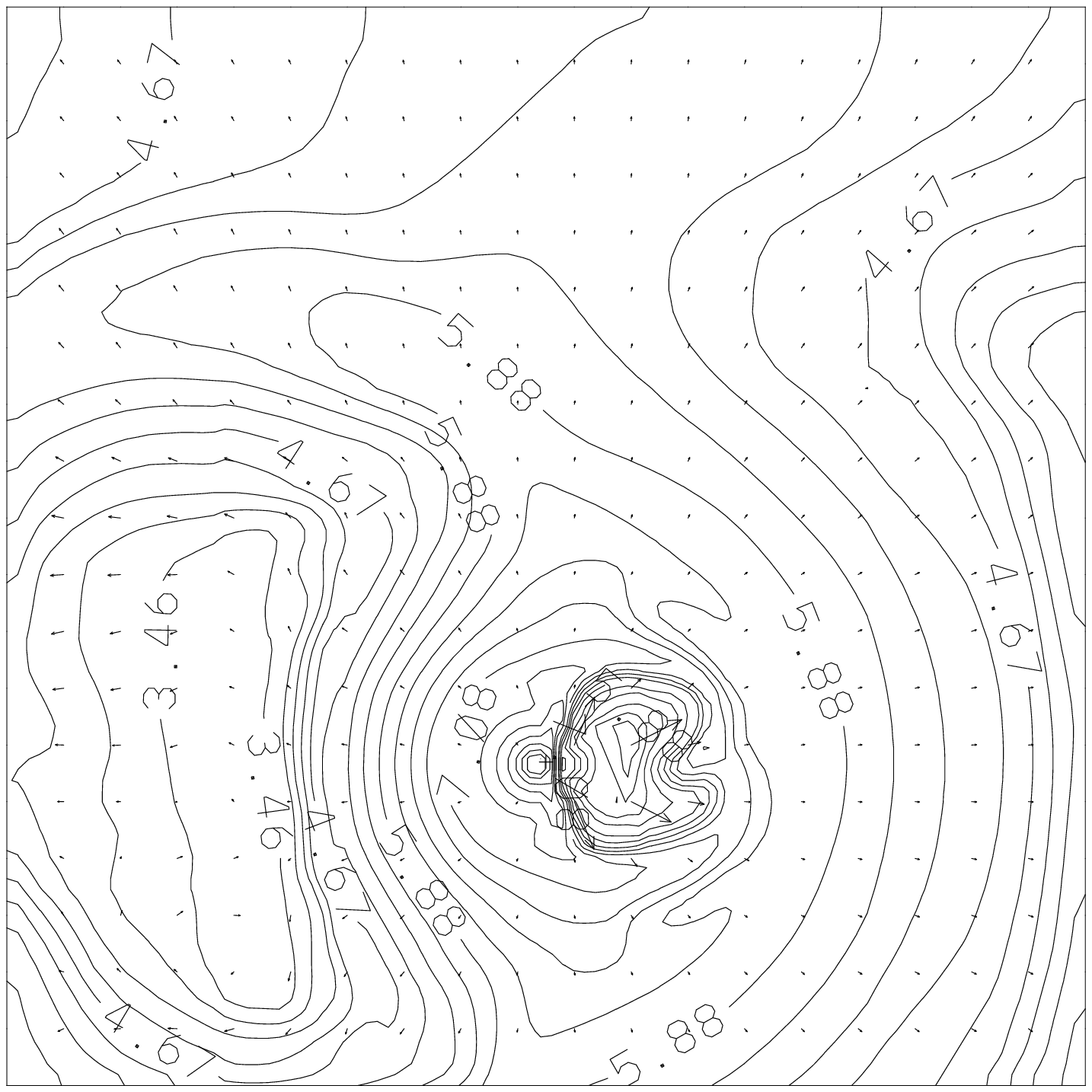}{4.4cm}{0}{30}{30}{-281}{-65}
   \plotfiddle{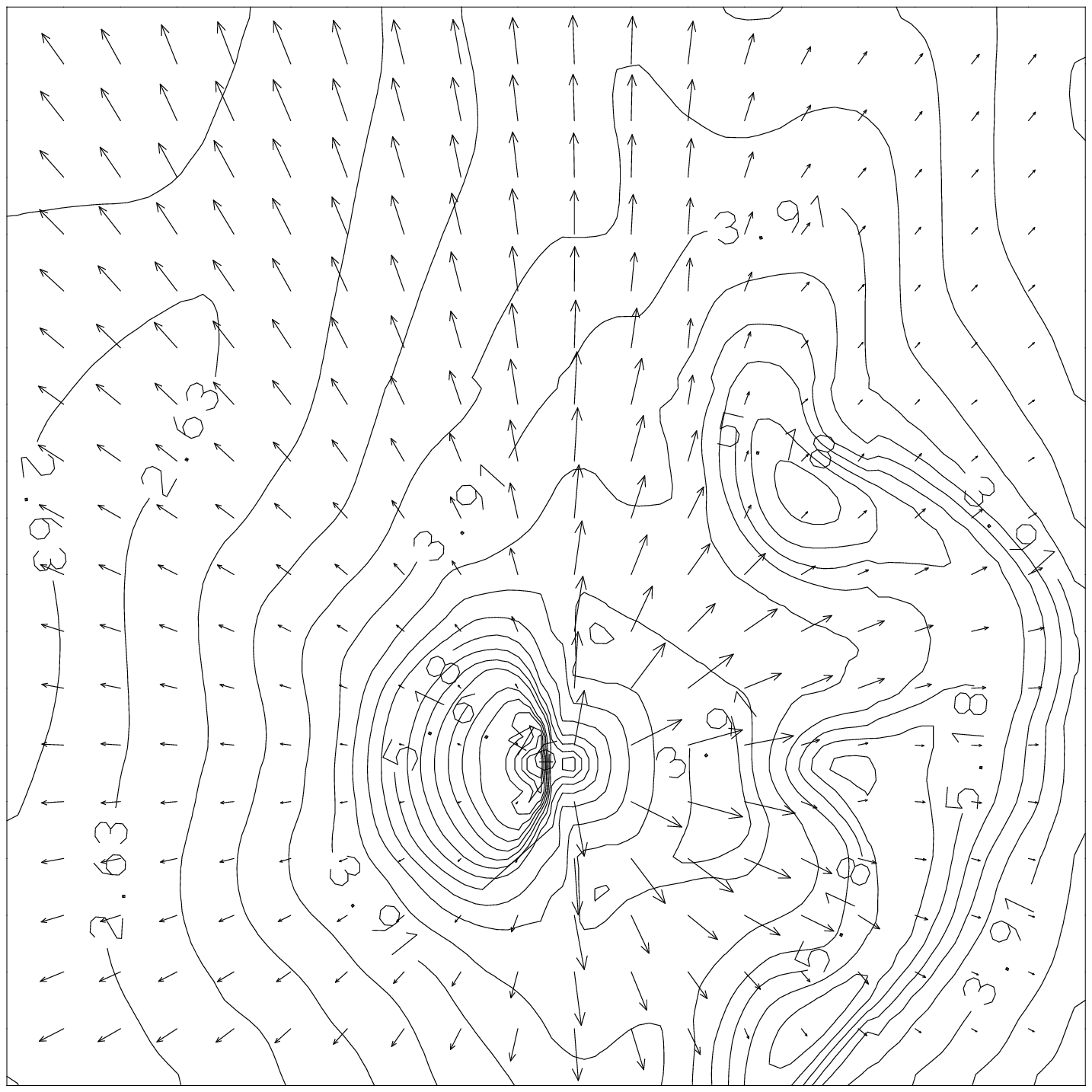}{4.4cm}{0}{30}{30}{-539}{-65}
           }
\caption{3D Density (logarithmically labeled) and projected velocity field
        ($v_{max} = 1000$~km/s) of the circumstellar matter for models
         weak (left), medium (middle) and strong (right). Top: slices
         along the orbital plane. The stars, rotating counterclockwise,
         are at the tip of the low-density cavity (hot) and in the
         center of the high-density region (cool). Bottom: slices normal
         to the orbital plane and along a line connecting the two
         stars. The stellar separation is $3\times10^{13}$~cm, the
         computational domain $1\cdot 10^{15}$~cm cubed.}
\label{fig:collwind_orbit}
\end{figure}

As the hot star works its way through the red giant wind, its own wind
pushes material aside, leaving behind a spirally shaped, low density
cavity. In the orbital plane as shown in
Figure~\ref{fig:collwind_orbit}, its role as a {\it rotating
snow-plow} becomes particularly apparent. The opening angle of the
spiral depends on the ratio of the momentum fluxes and is small in
model {\it weak} and large in model {\it strong}. Matter is piled up
at the leading edge of the spiral, where a shock bounded high density
shell confines the interaction zone (starting in the upper half in the
pictures). The trailing edge of the spirally shaped interaction zone
is characterized by a huge rarefaction wave, connecting the
high-density red star wind with the low density cavity of the fast
wind in a smooth, however steep way. As the temperature is
approximately constant across this trailing edge, the red-giant wind
is re-accelerated by the resulting pressure-gradient. The models,
therefore, predict a significant part of the red star wind to be
faster than single star winds. The leading and trailing part of the
interaction zone are connected by a small zone in the center where the
two winds collide head-on.

Looking now at the lower row of Figure~\ref{fig:collwind_orbit} we
notice in models {\it weak} and {\it medium} that the low-density
spirally shaped tube carved by the hot star grows in diameter with
increasing distance. The tube is eventually filled again when the
kinetic pressure of the fast wind is exhausted. However, this will
happen on a scale significantly larger than our computational box of
$1\cdot 10^{15}$~cm cubed. In both models the fast wind is embedded in
the red giant wind material. In model {\it strong} the situation is
different. Here the red star wind is basically restricted to a wide,
open, high-density spiral. Normal to the orbital plane, the dense red
giant wind is pushed away by the fast wind (see Section~\ref{sec:pne}).

We conclude that colliding winds force an extreme re-shaping of the
circumstellar material. The red star wind is no longer spherically
symmetric, nor is it smooth. Both, thin high density shells and huge
voids can be found even in the immediate neighborhood of the stars.

Finally, we note that the interaction zone of the colliding winds is
inherently unstable. High density knots and filaments are formed. For
a further discussion we refer to our contribution on colliding winds
in WR+O binaries (this volume), Walder \& Folini (1998a), and the
recent review of Walder \& Folini (1998b). Additionally, instabilities
induced by ionization may play an important role in symbiotics
(e.g. Garc\'{\i}a-Segura et al. 1999).
\subsection{The planetary nebulae link}
\label{sec:pne}
A common, bipolar, morphology and a similar dynamical behavior suggest
a link between some D-type symbiotics and planetary nebulae. We want
to make only two remarks here on this important link and refer to the
review of Corradi (this volume) comprehensive discussion.

The models discussed in Section~\ref{subsec:hydro_models} show that
the hot wind is confined by the dense wind from the red star whenever
its momentum flux is comparable to or weaker than that of the cold
wind. According to our simulations, the circumstellar material should
be strongly structured up to about 10$^{16}$~cm. However, since the
low-density voids occupy only a small volume, ionizing photons are
unlikely to penetrate that far and no large-scale optical nebula will
be present. But such a scenario may explain the radio measurements,
e.g. of AG~Peg by Kenny et al. (1991). On the other hand, if the
momentum flux of the fast wind is bigger, all material is blown away
in direction normal to the orbital plane, whereas in the orbital plane
the presence of the red star prevents an unhindered outflow.  A
bipolar-like structure is likely to extend to scales similar to those
of planetary nebulae. But even for a smaller momentum flux of the fast
wind a bipolar large-scale structure may form if the circumstellar
matter is more concentrated in the orbital plane, e.g. due to
accretion before outburst or due to rotation of the red star.


Due to previous or currently on-going wind accretion, symbiotic hot
stars have a good chance to be fast rotators and thus carry a larger
magnetic field than single white dwarfs. Thus, a new class of magnetic
wind models developed for planetary nebula may be of interest for
symbiotics. These models show (Chevalier \& Luo 1994; Rozyczka \&
Franco 1996; Garc\'{\i}a-Segura 1997) that rotating, magnetic CSPNe
with fields of some hundred Gauss can explain the observed variety of
shapes of planetary nebulae. In particular, elliptical and even
bipolar nebulae form quite naturally. In addition, due to magnetic
stresses, highly collimated jets can be formed (Garc\'{\i}a-Segura et
al. 1999). These jets have a very particular velocity law, where the
velocity is approximately linearly increasing along the
jet-axis. Exactly such a law was observed in the young planetary
nebulae MyCn~18 (Bryce et al. 1997). These results prove that the
presence of jets in a binary system does not require accretion.
\subsection{Spectral response}
\label{sec:spectra}
Spectra remain the main source of information on symbiotics although
imaging is becoming more and more important with the new generation of
telescopes. Due to the work of Corradi and Schwarz (e.g. Corradi \&
Schwarz 1993), we have fantastic images of the large scale structure
of D-type symbiotics. Radio images have also brought light into
nebular substructures of some S-type symbiotics.

There is a long tradition of applying photo-ionization codes to
symbiotic systems. Beginning with spherical symmetry (e.g. Nussbaumer
and Schild 1981), the models later were extended to axial-symmetry,
where the hot star as the ionizing source illuminates the spherically
symmetric wind from the cool star (e.g. Nussbaumer \& Vogel 1989;
Proga, Kenyon, \& Raymond 1998). Below we report on first attempts
where synthetic spectra are computed on the basis of 2D and 3D
hydrodynamical models, and thus include the influence of shocks, the
wind-wind interaction zone, and the orbital motion.

\subsubsection{Optical and UV}
Based on axi-symmetric hydrodynamical models, Nussbau\-mer \& Walder
(1993) investigated the influence of the wind-wind interaction zone on
the ionization structure and the spectrum of the symbiotic nebula.
Remarkably, the low-density cavity of the fast wind as well as the
high-density walls of the wind-wind-interaction zone are as important
as the luminosity and the temperature of the ionizing source. In
particular, the high-density shells of the interaction zone can block
high energy UV-photons, significantly reducing the emission area of
highly ionized species. Velocities in these shells are significantly
higher than red star wind velocities. Compared to spectra computed
from a single wind model, synthetic line profiles based on colliding
wind models show significantly broader feet. Lines of highly ionized
species are generally broader than lines of lower ionized species. In
addition, the line-shapes significantly depend on the line of sight
leading, suggesting that we should have a strong variation of the
profiles over an orbit of the system.

\begin{figure}[tbp]
%
\centerline{
   \plotfiddle{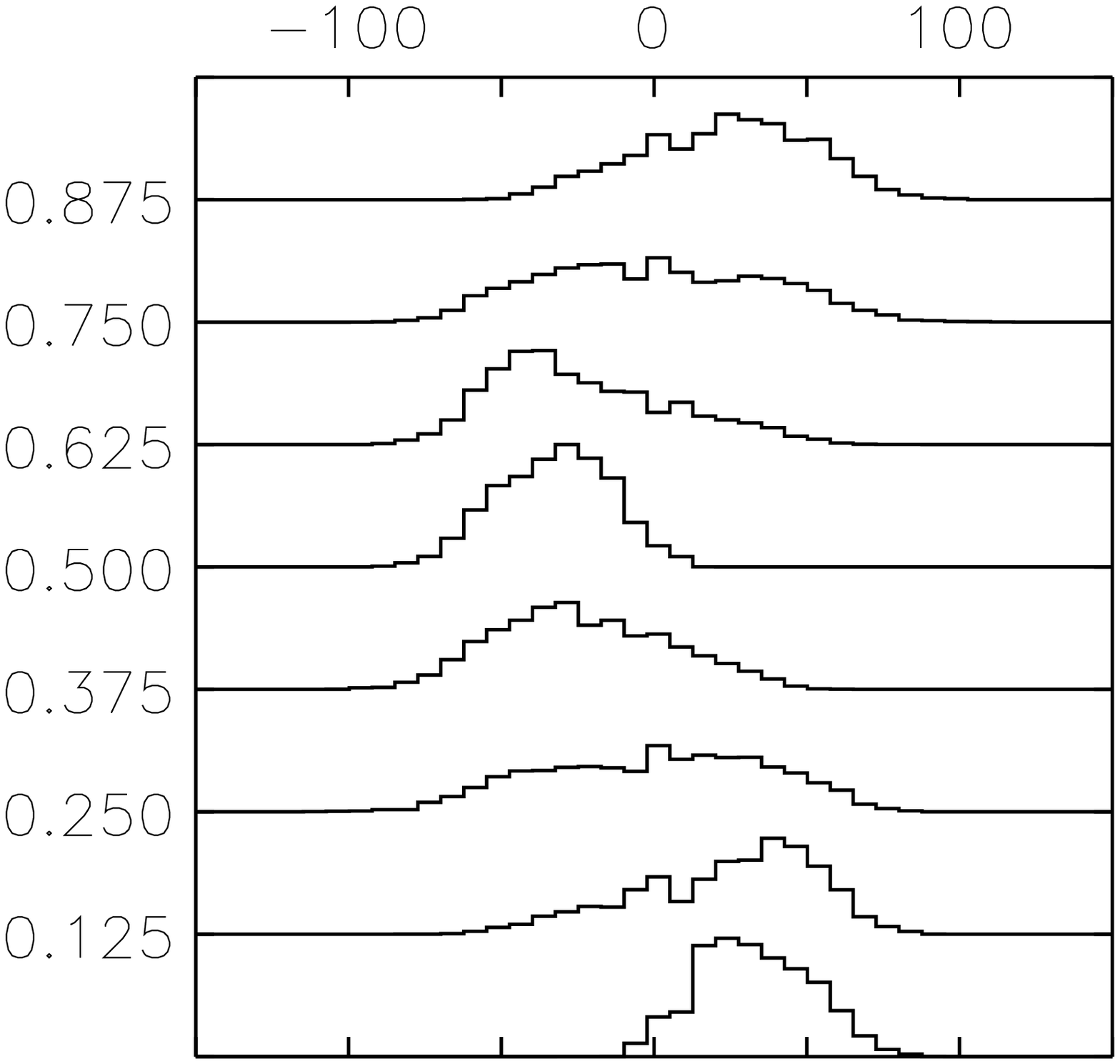}{4.1cm}{0}{28}{28}{-10}{-51}
   \plotfiddle{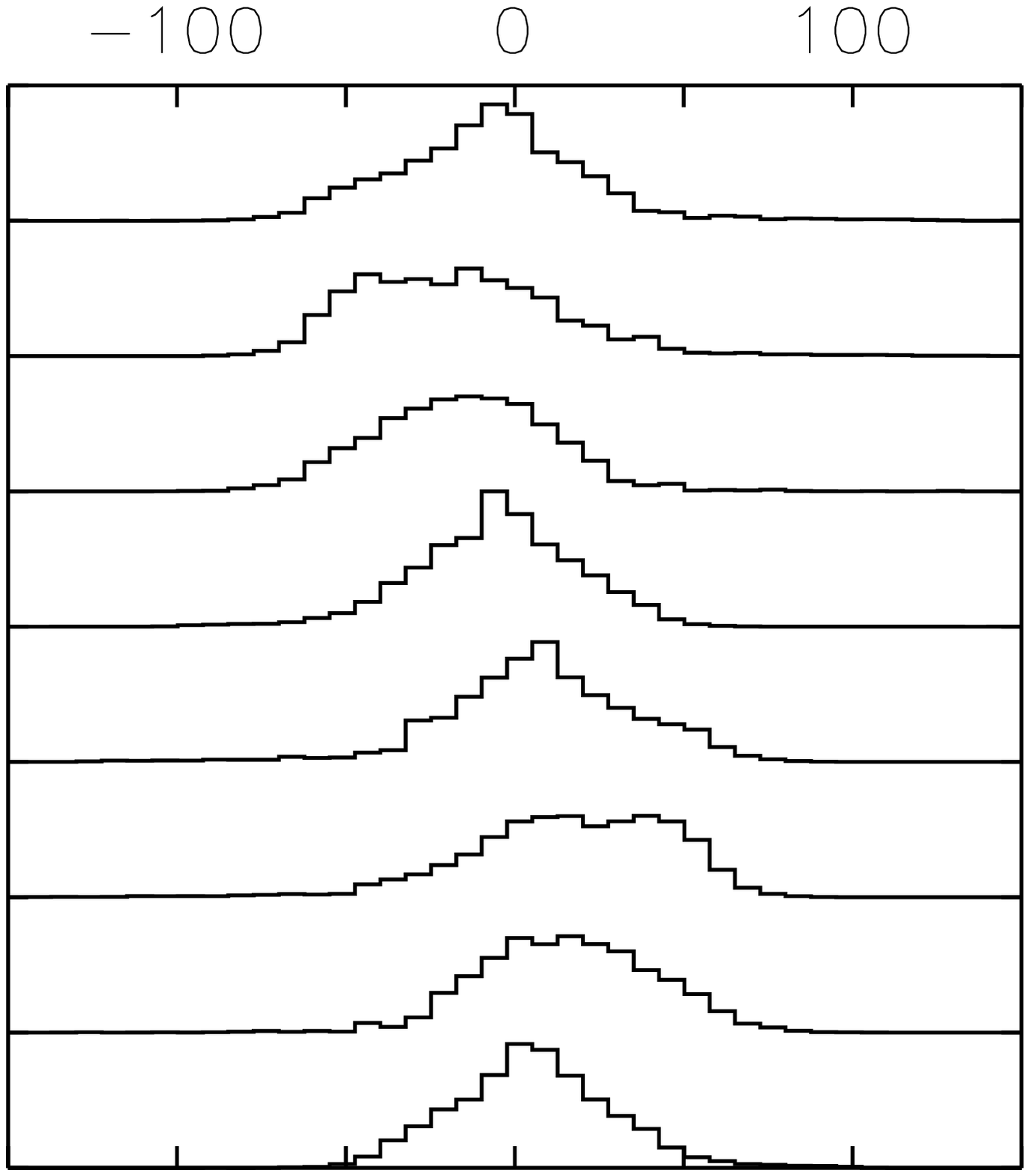}{4.1cm}{0}{28}{28}{-276}{-51}
   \plotfiddle{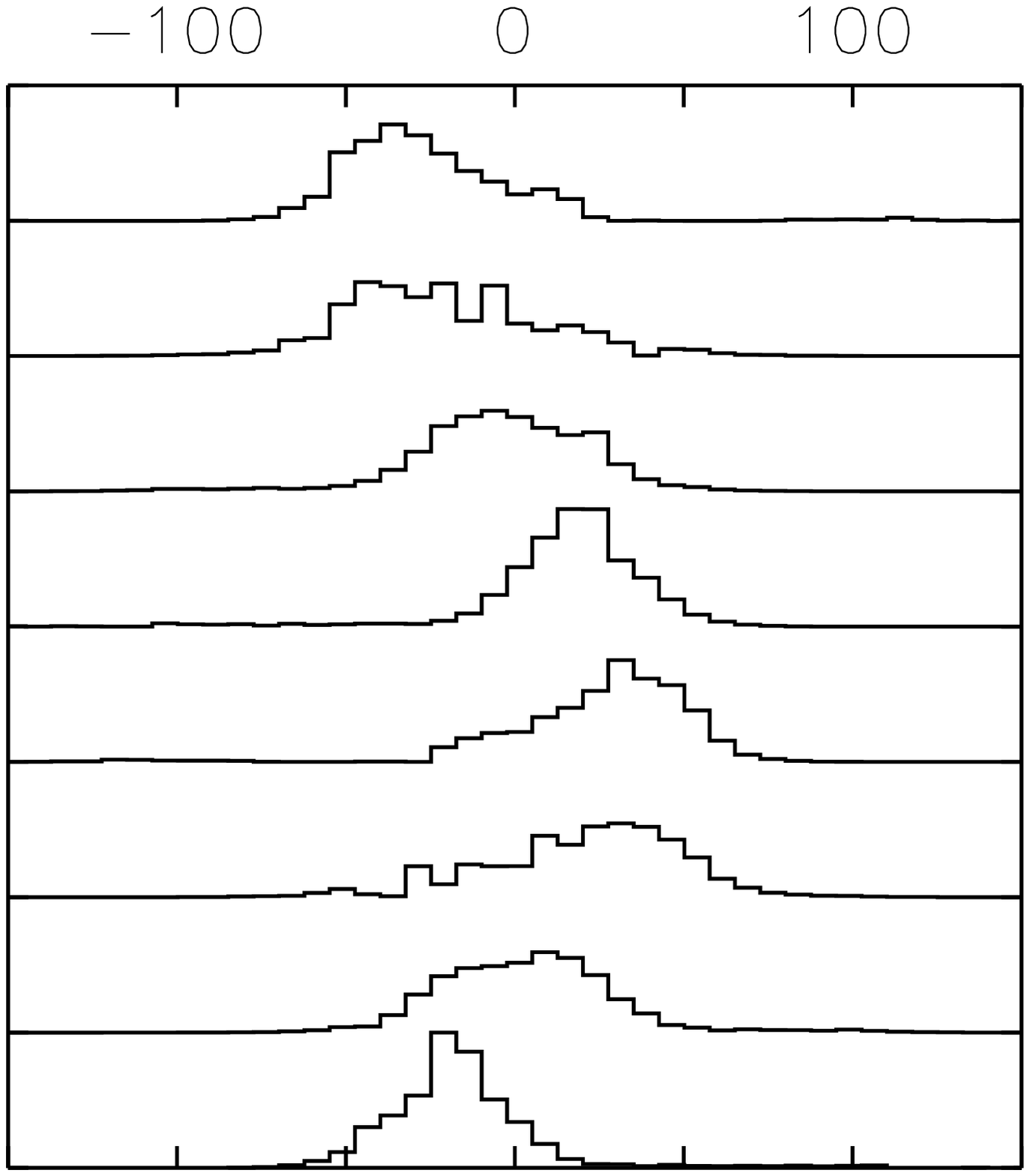}{4.1cm}{0}{28}{28}{-541}{-51}
   }
\caption{Phase-dependence of synthetic line profile for
        [O{\sc iii}] $\lambda5008$ for the models 
        weak (left), medium (middle), and strong (right) with
        a hot star of 100 L$_{\sun}$ and 90'000~K. Profiles 
        are computed for an observer in the orbital plane, 
        phases are increasing from bottom to 
        top, starting at phase zero (eclipse of hot star). 
        Each line profile is plotted against velocity in km/s, negative
        velocities corresponding to motion towards the observer.}
\label{fig:collwind_spectra}
\end{figure}
This was confirmed by the work of Folini (1998) who computed orbital
variations of synthetic line profiles on the basis of the
hydrodynamical models presented in Section~3.1. It was also shown
there that the line profiles as a whole are shifted as a function of
orbit. Moreover, it was demonstrated (see
Figure~\ref{fig:collwind_spectra}) that the same line ([O{\sc
iii}]$\lambda5008$) computed for the same ionizing source (100
L$_{\sun}$ and 90'000~K) may be shifted completely differently in each
of the three models. If the line emission stems primarily from the
immediate vicinity of the cool companion the line profile shows a
maximum blue shift around phase 0. On the other hand, if the emission
stems from the interaction zone, maximum blue shift is reached around
phase 0.5. It was further shown that for each of the investigated
models and ionizing sources it is possible to find ions probing the
interaction zone.

\subsubsection{X-ray emission}
Observed strong X-ray emission stood at the beginning of colliding
wind models of symbiotics. For the symbiotic novae HM~Sge Kwok \&
Purton (1979) suggested a model where a shocked fast wind from the hot
star leads to a spherical hot bubble. In contrast, Wallerstein et
al. (1984) and Willson et al. (1984) suggested the observed X-rays to
come from the head-on collision of the fast wind from the hot star
with the slow, dense wind from the red star. Present models discussed
in Section~3.1 suggest something in between. The
collision zone of the two winds, and therefore the distribution of the
hot plasma, is spirally shaped. Additionally, the hot star is expected
to contribute with a supersoft component to the X-ray spectrum. First
computational models of such two-component spectra were published by
Walder \& Vogel (1993) and M\"{u}rset, Jordan \& Walder (1995).

{\bf Observations:} In a systematic study of 16 symbiotics M\"urset,
Wolff, \& Jordan (1997) detected 60 percent of them as X-ray
sources. All of them but two accreting neutron stars have a supersoft
component.  Seven show emission of an optically thin plasma with
temperatures between 3-20 million Kelvins and the authors suggest an
observational relation of $L_{\mbox{\small hot plasma}} = 10^{-5}
L_{\odot}$.

{\bf Theory:} The case of a spirally shaped interaction zone described
in Section~3.1 causes the entire fast wind to become shocked, at least
for cases where the momentum flux of the red star wind dominates that
of the fast wind. The resulting X-ray luminosity can be estimated as
\begin{equation} 
L_{xray} = x_{eff} \times L_{mech}
         \stackrel{CAK}{\approx} x_{eff} \times 10^{-5}
           \left(L_{*} / L_{\odot} \right)^{1.5}
            L_{\odot},
\end{equation}
where the efficiency factor $x_{eff}$ lies between one and ten
percent, depending on model parameters. So $L_{xray}$ is for luminous
stars probably higher than the observed $L_{\mbox{\small hot plasma}}$
but the observational error bars are still large and in the
theoretical prediction for $L_{xray}$ circumstellar absorption has
been neglected. Concerning temperatures, the highest temperatures are
reached in the system center where the flows collide head on. However,
the bulk of the fast wind hits the spiraling interaction zone at
angles much smaller than 90 degrees and the temperatures reached are
smaller.

{\bf Discussion:} Assuming a colliding wind scenario, the X-ray
emission is closely linked to the wind parameters. However, X-ray
observations of EG~And and AG~Peg seem inconsistent with their wind
parameters derived from observations. CAK fit badly as well, but here
the situation may be saved.

Using a one temperature fit to the observed X-ray spectrum M\"urset et
al.  (1997) find $1.5\cdot 10^{7}$~K for EG~And. The wind parameters
derived by Vogel (1993) lead to a theoretical peak temperature of only
$5.6 \cdot 10^6$~K, CAK-based wind parameters result in unrealistic
$1.5 \cdot 10^9$~K. Nevertheless, CAK-winds are not excluded by the
observed X-ray spectrum of EG~And for the following reasons. The bulk
emission seen in X-rays is cooler than the theoretical peak
temperature. Second, if heat conduction were taken into account peak
temperatures would be generally reduced. As shown for hot star
binaries by Myasnikov \& Zhekov (1998) and by Motamen, Walder \&
Folini (1999) heat conduction becomes important above about $10^7$~K
and will reduce the temperature by up to an order of magnitude. For AG
Peg the situation is similar but less pronounced. From X-ray
observations M\"urset et al. (1997) derive $3.16\cdot
10^6$~K. Observed wind velocities result in a peak temperature of $1.5
\cdot 10^7$~K, CAK based parameters lead to $4 \cdot 10^8$~K. While
both sets of wind parameters seem possible, the CAK values may be
preferable as again the X-ray observations reflect the bulk and not
the peak temperature, and as heat conduction reduces peak
temperatures. Finally, when taking observation based wind parameters
the efficiency factor $x_{eff}$ must be below 0.001 to fit the
observed X-ray flux. This is in contradiction with simulations.
Moreover, CAK wind values result in $x_{eff} \approx 0.1$, a value
which is more realistic.
\section{Accretion}
\label{sec:accretion}
According to current knowledge, the existence of every symbiotic
system requires accretion at some stage. Yet, some symbiotics are
observed to be colliding wind binaries. Is the observed symbiotic
phenomenon compatible with accretion as well? The alternative would be
that whenever accretion takes place the symbiotic signatures
vanish. On observational grounds, this question has not been
settled. And while colliding wind models are now compared to
observations, accretion models are not yet sufficiently evolved.
In the discussion below, most aspects are, however, of more general
nature.

{\bf Observations:} Accretion sets free a relatively small amount of
energy but various observational signatures may be explained in terms
of accretion. As discussed by M\"{u}rset et al. (1991), observed UV
nebular spectra from symbiotics require a more compact and hotter
ionizing source than the emission from a classical Keplerian accretion
disk. On the other hand, Sokoloski \& Bildsten (1999) argue that the
detected variation of 1682 seconds in the optical emission of Z~And
may be explained by accretion onto a highly magnetic white dwarf.
They attribute the outbursts to classical disk-instabilities of a
Keplerian disk. Looking at X-ray observations (temperature,
luminosity, time variability), CH~Cyg seems to behave like a CV
(Ezuka, Ishida \& Makino 1998, based on ASCA-spectra). However,
according to M\"{u}rset et al. (1997) its X-ray properties are
different from any other symbiotic system, which all shows
significantly lower temperatures that may be explained in terms of
colliding winds. However, all these latter observations are ROSAT data
only, therefore lacking a high-energy channel.

{\bf The case of RW Hya:} For RW Hya there are indications that it is
a wind accreting system. If true, RW Hya is the first confirmed
accreting symbiotic system. Dumm et al. (2000) discovered an
unexpected occultation of the hot component at phase $\phi=0.78$. This
occultation is unrelated to the eclipse of the hot component. The
occultation lasts approximately $\Delta \phi = 0.04$. The spectral
characteristics of this event indicate Rayleigh scattering due to a
high column density of neutral hydrogen in the line of sight to the
hot star. The authors interpret this observation in terms of an
accretion wake filled with highly compressed material, trailing the
white dwarf. They corroborate this suggestion with hydrodynamical
simulations which show the formation of such a wake at approximately
the correct orientation and opening angle.

{\bf Theory:} Wind accretion in separated but heavily interacting
binaries with slow winds is not yet well understood. In S-type
symbiotics Bondi-Hoyle-Lyttleton theory is not valid since here the
Bondi accretion radius is comparable to the stellar separation. Walder
(1997) reports that in such a situation only 6 percent of the formal
Bondi-Hoyle value can be accreted, corresponding, however, to 6
percent of the mass loss rate of the secondary. For binaries in which
the Bondi accretion radius is small compared to the stellar separation
(e.g. HMXRB) 63 percent of the Bondi-Hoyle accretion rate is reached,
corresponding, however, to only 0.6 percent of the mass loss rate of
the secondary. For D-type symbiotics, where the separation is a factor
of 10-20 larger, Bondi-Hoyle-Lyttleton theory may be applicable but we
are not aware of any simulations.

Hydrodynamical studies of accreting systems with dynamical parameters
comparable to S-type symbiotics were performed by Theuns \& Jorissen
(1993), Bisikalo et al. (1995, 1996), Theuns, Boffin, \& Jorissen
(1996) and Mastrodemos \& Morris (1998). Although there are
significant differences between their results (see next paragraph) and
despite their insufficient resolution close to the accreting star, all
models agree on some issues: 1) A large fraction of the donor-wind is
captured by the accretor (up to 10 percent). 2) In the vicinity of the
accretor the density is strongly enhanced in the orbital plane and the
flow spins. Some authors call this structure a disk. But even though
the flow is spinning, it is still advection dominated and far from the
regime of a viscous, Keplerian disk. Strong shocks are visible. 3)
There is spin up of the accreting star. 4) All show complexly shaped
nebulae on a scale of a few stellar separation. The wind from the red
star is far from being spherically symmetric.

The models, however, differ in one important aspect. According to
Bisikalo et al. (1995, 1996), a very extended (more than
70~R$_{\odot}$!) disk is formed with no sign of a wake. All other
results show a much smaller spinning structure with a very prominent
wake trailing the accreting star. The main difference between these
two simulation and all others is that Bisikalo et al. apply a
Roche-potential based on both stars and neglect forces which
accelerate the wind from the red star, whereas the other simulations
all assume a net accelerating force (driving forces overwhelming
gravity) from the donor star, together with gravitation from the
accretor. None of the models considers radiative forces from the
accreting star which, in fact, could be quite large as discussed in
Section~2.

Bisikalo et al. (1996) provide synthetic H$\beta$-profiles on the
basis of their 2D hydrodynamical simulations. On top of a very broad
foot -- emitted by the disk -- a thin nebular line can be found,
varying in shape over an orbit. We know of no comparison of these
profiles with observations. 

We conclude that accretion models predict a highly aspherical
distribution of the circumstellar matter with a clear concentration
in the orbital plane around the accreting star. Presumably, the
disk-like structure is optically thick. Its spectrum can, however,
probably not be compared with that of a Kepler disk since the flow
is still advection dominated.
\section{Summary}
\label{sec:conclusion}
To explain the observed number of symbiotic systems, consisting of a
hot post-AGB or pre-white dwarf and an evolved low-mass star,
accretion must occur at some stage. So far, there is no direct
observational proof that accretion takes place in any system we
classify as symbiotic. For colliding winds, on the other hand, such
evidence most likely exists for at least one system. In fact, the
question whether accretion can occur at all in a system we observe as
symbiotic is still under debate, as is the question of how the
accretion takes place. A classic Keplerian disc is not mandatory.

One of the key questions for both, colliding winds and accretion, is
how the matter close to the white dwarf or in its atmosphere responds
to the radiation field of this star. The overwhelming majority of
symbiotics would have to be colliding wind systems if CAK theory were
applicable for the wind of the white dwarf. However, there are
inconsistencies between CAK-theory and observations which are not
understood up to now. Likewise it has barely been investigated how
infalling, accreting matter would be affected by the radiation field
of the accretor. Could radiation pressure in continuum and lines
prevent accretion?

In both, colliding wind and accretion systems, the circumstellar
matter -- consisting of the wind of the red star -- is highly
structured and by far not spherically symmetric. This will have severe
consequences for the nebular spectrum. In all spectral ranges,
comparison with observational data is more advanced for colliding wind
models than for accretion models. Presently, neither of the two models
can be rejected on this basis. In particular, features like jets and
bipolarity can also be explained in the frame of colliding wind
models.

Comparing typical lifetimes of AGB or RGB stars with typical times for
accretion and subsequent shell flashes or novae suggests that several
accretion phases can take place during the life of the AGB or RGB
star. The primary components in symbiotics then would be alternatingly
accreting and wind-shedding stars. Whether the symbiotic phenomenon is
observable during both phases is not yet clear. Also the possibility
that accretion is accompanied by outflow can not be ruled out. A
strict division between colliding winds and accretion then might not
be possible. Other exciting years of theoretical and observational
research lie ahead until we have understood symbiotics, stellar
systems among the most complex ones.

\acknowledgments

The authors benefited from fruitful discussions with their colleagues
Harry Nussbaumer, Hansruedi Schild, Hans-Martin Schmid, and in
particular with their room-mates Thomas Dumm and Urs M\"urset.

\end{document}